# RobustPdM: Designing Robust Predictive Maintenance against Adversarial Attacks

Ripan Kumar Kundu[a], Ayesha Siddique[a], Gautam Raj Mode[a], Khaza Anuarul hoque[a,*]

[a]*Department of Electrical Engineering & Computer Science, University of Missouri, Columbia, 65201, MO, USA*

**Abstract**

The latest developments in Industry 4.0 are driving the evolution of smart manufacturing environments, with increased use of Internet-of-Things (IoT) and Deep Learning (DL). Predictive maintenance (PdM) techniques have proven to be highly effective in reducing maintenance expenses and downtime for complex machinery, while also enhancing overall productivity in industries through data-driven decision-making. These modern PdM systems employ IoT devices data and DL algorithms to predict Remaining Useful Life (RUL). Unfortunately, IoT sensors and DL algorithms are both prone to cyber-attacks. For instance, DL algorithms are known for their susceptibility to adversarial examples. Such adversarial attacks have been extensively studied in computer vision and related domains. However, it is surprising that their impact on the PdM domain is vastly under-explored. This is because the adversarial attacks in the computer vision domain for classification tasks cannot be directly applied to the PdM domain for multivariate time series (MTS) regression tasks. Thus, modern data-driven intelligent PdM systems pose a significant threat to the safety and cost-critical applications. In this work, we propose an end-to-end methodology to design adversarially robust PdM systems by extensively analyzing the effect of different types of adversarial attacks and proposing a novel adversarial defense technique for DL-enabled PdM models. First, we propose novel MTS Projected Gradient Descent (PGD) and MTS PGD with random restarts (PGD_r) attacks. Then, we evaluate the impact of MTS PGD and PGD_r along with MTS Fast Gradient Sign Method (FGSM) and MTS Basic Iterative Method (BIM) on Long Short-Term Memory (LSTM), Gated Recurrent Unit (GRU), Convolutional Neural Network (CNN), and Bi-directional LSTM based PdM system. Our results using NASA's turbofan engine dataset show that adversarial attacks can cause a severe defect (up to 11X) in the RUL prediction, outperforming the effectiveness of the state-of-the-art PdM attacks by 3X. Furthermore, we present a novel approximate adversarial training method as a defense against adversarial attacks. We observe that approximate adversarial training can significantly improve the robustness of PdM models (up to 54X) and outperforms the state-of-the-art PdM defense methods by offering 3X more robustness.

## 1. Introduction

Recent trends in Industry 4.0 have paved a path for evolution in smart manufacturing environments [36], including extensive utilization of Internet-of-Things (IoT) and Deep Learning (DL). In particular, the DL and IoT sensors have enabled the data-driven *predictive maintenance (PdM)* [22] in achieving a competitive edge in Industry 4.0 by preventing asset failure, analyzing production data, and making data-driven informed decisions to predict issues before they happen. This indeed results in reduced maintenance cost, downtime, and increased productivity. The IoT sensors can sense changes in the physical world (such as temperature, vibration, pressure, etc.) and analyze the sensed data using the state-of-the-art DL algorithms, which help to ascertain the Remaining Useful Life (RUL) of equipment in use [58]. This enables a highly reliable and cost-efficient industrial automation framework. Unfortunately, IoT sensors are susceptible to cyber attacks [45], and DL algorithms can also be fooled by carefully crafted adversarial examples [53]. According to a recent survey, a 50% increase in the daily average number of cyber attacks was observed in the third quarter of 2022 compared to its first half. The state-of-the-art PdM systems that utilize IoT devices and DL algorithms tend to inherit such vulnerabilities; hence, they are a lucrative target for cyber-attackers.

In a large-scale industrial environment, adversarial attacks on the PdM system can lead to a wrong maintenance decision, e.g., an incorrect prediction of RUL can result in delayed maintenance and unexpected failures. Such unexpected failures are considered a primary operational risk, as they can hinder productivity and incur a huge loss. For example, industrial batteries find their applications in telecommunications, uninterruptible power supplies, systems for safety/alarms, asset tracking systems, and medical equipment. In the use case of batteries in telecommunications, an incorrect prediction of the RUL of the battery may result in the loss of critical information. In another situation, a wrong prognostic prediction in an autonomous vehicle or aircraft may lead to the loss of human lives. Indeed, IoT and ML are revolutionizing the prognostic domain; however, their security vulnerabilities (related to IoT sensors and ML algorithms) pose a great challenge for Industry 4.0.

*Limitations of the existing works and contribution of this paper.* Integrating IoT sensors and DL in PdM systems cre-ate an interdisciplinary and novel *adversarial predictive maintenance* problem. Indeed, an adversary can launch network-based attacks, such as denial of service (DoS), attack and man in the middle attacks, Replay attacks [20, 21] etc., to disrupt the PdM systems. However, compared to network-based attacks, adversarial attacks add very small perturbations to the IoT sensor readings, and the perturbed readings often lie in the bound of the sensor's expected readings. Thus, such attacks are extremely stealthy, and it is very hard to detect them. In the past, the impact of adversarial attacks has been extensively studied in the computer vision domain [3] and also has been explored in other domains such as audio [19], image, graph, and text [61, 50], typically for classification applications. Traditionally, the main concentration of designing PdM systems is their accuracy of prediction while ignoring their robustness. It is indeed quite surprising that the impact of adversarial attacks in this domain is vastly under-explored. In fact, our previous works in [33, 34] are the first two works in the PdM domain that reported the vulnerability of PdM systems against adversarial attacks and network-based attacks, respectively. Specifically, our previous work in [33] was the first work to modify the Fast Gradient Sign Method (FGSM) [16], Basic Iterative Method (BIM) [25] attacks for multivariate time-series regression (MTSR) models and then, apply them to PdM applications leading to erroneous RUL calculation. However, one of the significant limitations of [33] is that it only investigates the impact of adversarial attacks and does not propose any effective defense mechanisms that can be deployed for PdM systems. Recently, in [6], the authors reported the impact of universal adversarial attacks on PdM models. In another work, [18] the impact of *white-box* attacks on PdM models is studied, and a stacking ensemble learning-based framework is reported to be adversarially robust. Note, in *white-box attacks*, the adversary knows the model, including training data and model parameters, while in *black-box attacks*, the adversary lacks this information. However, the adversary can still cause significant damage by using the *transferablity* property of adversarial examples, as they can often transfer from one model to another. This means that it is possible to attack models to which the attacker cannot access [53, 29]. Such a scenario was ignored in [18, 6] as they did not explore the impact of black-box attacks via attack transferability analysis. Note, analyzing the impact of various perturbation budget $\epsilon$ is a standard process for evaluating the effectiveness of an attack on DL models. This is because the larger the value of $\epsilon$ is, the higher is the impact of an attack. However, choosing a large value of $\epsilon$ makes an attack easily detectable. Thus, an attacker should always choose an $\epsilon$ value that is not too large but enough to cause significant damage. Unfortunately, an extensive analysis with different values of perturbation budget $\epsilon$ is also ignored in [18]. Indeed, the PdM domain still lacks an end-to-end methodology for making the PdM models robust and limiting the transferability of both black-box and white-box attacks with an effective defense. Motivated by this, in this paper we present:

1. A novel design exploration methodology (**RobustPdM**) for extensively analyzing the accuracy and developing a robust DL-enabled PdM system under a different types of white-box and black-box adversarial attacks. The proposed RobustPdM methodology can help PdM designers not only to analyze the accuracy of the PdM models but also their robustness.

2. The formalization of the adversarial predictive maintenance problem and two new adversarial example generation algorithms (in addition to the two attacks presented in [33]) for PdM applications.

3. A novel defense mechanism *approximate adversarial training* for PdM systems under adversarial attacks. We also compare its effectiveness in defending PdM systems to the traditional non-approximate adversarial training technique, which is popular in the computer vision domain [24].

4. An empirical case study using NASA's C-MAPSS dataset [42] to study the quantitative adversarial robustness of PdM systems.

---

*Corresponding author

It is worth mentioning that adversarial attacks from the computer vision domain cannot be directly applied to the PdM domain due to the nature of the attacks (pixels vs. time series, classification vs. regression). Fundamentally, the regression tasks (which is the case for PdM) use mean squared error (MSE) loss in contrast to the classification problems that use the cross-entropy loss to minimize the expected value of the cost function in the training phase. Moreover, in computer vision, small perturbations are added to the images' pixels, leading to the misclassification of the images. In contrast, in the PdM domain, the attackers' objective is to add small perturbations to IoT sensor readings leading to erroneous RUL prediction. In this paper, the algorithms proposed for crafting the adversarial examples are inspired by the methods that are popular in the computer vision domain, i.e., Fast Gradient Sign Method (FGSM) [16], Basic Iterative Method (BIM) [25], Projected Gradient Descent (PGD) [30] and PGD with random restarts (PGD_r). We formalize these attacks for MTSR models and then apply them to craft adversarial examples for Long Short-Term Memory (LSTM), Gated Recurrent Unit (GRU), Convolutional Neural Network (CNN), and Bi-directional Long Short-Term Memory (Bi-LSTM) based PdM models as they are known for their tremendous success in the prognostics domain [63, 65, 39, 27, 44, 8]. We analyze both, the *white-box* and *black-box* attack scenarios. Our analysis reveals that adversarial examples can seriously defect remaining useful life (RUL) predictions. Specifically, PGD and PGD_r can defect RUL up to 3X when compared to the state-of-the-art robust optimization method (ROM) and momentum iterative method (MIM) attacks in [18] and up to 4X when compared to the FGSM, and BIM attacks in [33]. Furthermore, we present a novel PdM adversarial attack defense mechanism *approximate adversarial training*, which applies quadratic coefficients through approximation of model weights and averages the loss gradient over the batches of adversarial examples. We also compare the effectiveness of the proposed defense mechanism against conventional non-approximate adversarial training that is widely used in the computer vision domain (interestingly, never been applied to the PdM domain).

Our results show that the proposed approximate adversarial training can improve the robustness of PdM models upto 54X, i.e., 11X higher than the conventional non-approximate adversarial training approach and 3X higher than the state-of-art methods [18]. This paper builds upon our prior results in [33] and significantly extends by proposing the novel RobustPdM methodology, two new attacks (MTS PGD and MTS PGD_r), analyzing attacks on a new PdM model (Bi-LSTM PdM), and by proposing novel adversarial defense methods (approximate adversarial training) in the PdM domain with new robustness results.

*Paper organization.* The rest of the paper is organized as follows. Section II provides the preliminary information about IoT sensors, DL enabled PdM, and their vulnerabilities. Section III presents the methodology for analyzing and mitigating the impact of adversarial attacks on PdM systems. Section IV formalizes the robustness and presents the algorithms for crafting the adversarial examples for MTS FGSM, MTS BIM, MTS PGD, and MTS PGD_r attacks, and approximate adversarial defense in PdM systems. Section V presents their corresponding experimental results with a case study. Finally, Section VI concludes the paper.

## 2. Background

In this section, we present a brief overview of the IoT sensors and DL algorithms in PdM systems and their adversarial vulnerabilities.

### 2.1. IoT and DL in PdM

Fig. 1 shows an overview of the cloud and edge centric PdM architecture that employs both IoT sensors and DL algorithms. It is comprised of four layers: physical, edge, cloud, and visualization layers [2]. The *physical layer* collects sensor measurements (e.g., temperature, pressure, etc.) from different industrial equipment using the IoT sensors [4]. The acquired signals are then sent to the edge layer for data processing. The *edge layer* pre-process the incoming data to make it useful for equipment monitoring process. The state-of-the-art PdM systems employ equipment monitoring, which is an automated approach that eliminates the need for manual inspection/analysis. It uses the parameter RUL for indicating the amount of time left before a piece of equipment or machine fails or degrades to a point at which it cannot perform its intended function anymore. The RUL is required to be estimated accurately to avoid both early and late prediction, which may lead to over-maintenance and catastrophic failures. Therefore, the edge layers in an edge centric PdM systems use DL as a complementary data-driven approach. The DL is proven to be effective in analyzing the immense volume of pre-processed sensing data for accurate RUL insights [64]. The state-of-the-art DL algorithms [59, 48, 15], especially LSTM, GRU, Bi-LSTM, and CNN, have shown great success in such prognostics tasks [63, 65, 39, 27, 44, 8]. After completing these prognostics, the resulting data is transmitted to the cloud servers. *The cloud layer* performs in-depth data analysis and also, trains/re-trains the DL models for the newly arrived edge data. After the re-training, the re-trained DL model is sent back to the edge layer for precise state prediction of the equipment. The *visualization layer* uses the data collected from the field, along with the results from the DL model, for providing a visual representation of the actionable insights to the end user.

### 2.2. Vulnerabilities in PdM systems
#### 2.2.1. Attacks on IoT sensors

The recent advent of IoT devices and deep learning has led to an anticipated adoption of smart analytics in cyber-physical domain. The cyber-physical systems are undoubtedly vulnerable to multiple threats and thereby, also to new means of attacks on IoT sensors [47, 12]. The data collected by IoT sensors has a great impact on the decisions related to prognostics and mainatainence in PdM systems. Even if the network is secured, an attack on the IoT sensor leads to incorrect decisions and may cause a considerable performance loss in the PdM systems [1]. For example, an attacker can utilize sensors for transfering a malicious code, triggering the messages to activate a malware planted in an IoT device [52], capture sensitive information shared between devices [31], or even capturing encryption and depreciation keys to extract the encrypted information [10]. This can be better explained using Figure 1. An adversary can target the physical sensors and also, the communication channel between the IoT sensors and edge devices. An adversary can access the sensors data to craft the adversarial examples and inject them into the PdM systems as a False Data Injection (FDI) [41] attack. Indeed, the cloud layer also has its vulnerabilities [23]; however, third-party cloud services, such as AWS, Azure, etc. have typically their security measures [5] which makes them less vulnerable when compared to the physical and edge layer.

Understanding the above discussed IoT sensor-based threats have gained great interest in the recent years from both academia and industry. This help the researchers and industrialists in detecting and preventing them efficiently. IoT sensors are small battery-powered or energy harvesting devices and hence, limited by power or energy constraints. These limited resources are indeed an obstacle for employing the most state-of-the-art complex security mechanisms for detecting the security threats on IoT sensors [51]. Note, detection and defense for IoT sensor-based threats is also an active area of research in the cyber physical domain [11]. However, how such IoT-based sensor threats influence ML algorithms in making the incorrect decisions for industrial automation is yet to be explored extensively.

#### 2.2.2. Adversarial attacks and defense in DL-based PdM

Besides IoT attacks, the performance of a PdM system can also be compromised by the orchestrated security attacks on DL algorithms. The study of the effect of adversarial attacks on machine learning techniques is known as *adversarial machine learning*, which is one of the most active research topics in the deep learning community [53, 25]. However, when it comes to PdM domain, there are only afew works that study adversarial degradation in PdM models. Recently, in [6], the authors studied the impact of universal adversarial attacks on the multivariate time series DL based regression models. In another work,

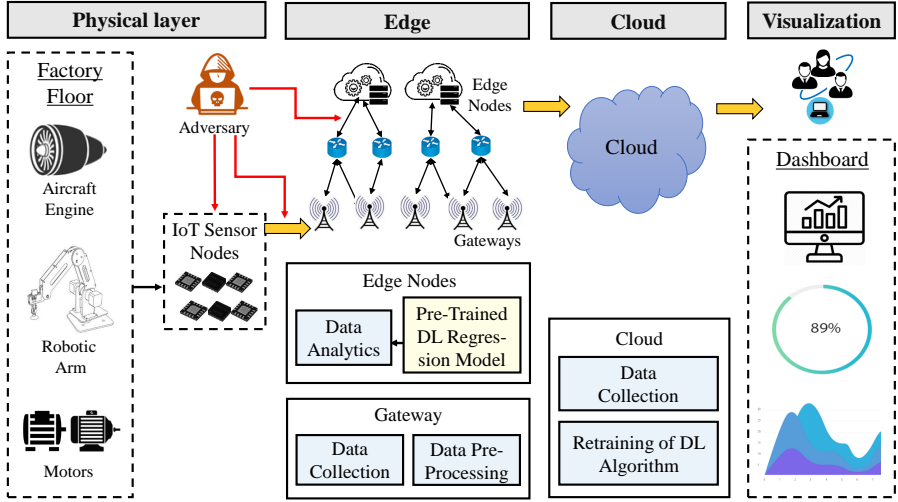

Figure 1: PdM cloud-edge architecture and threat model

[18], the impact of *white-box* attacks on different PdM models is studied. Nevertheless, the black-box attacks are more perturbing and realistic when compared to white-box attacks. The impact of black-box attacks is yet not investigated especially, by using the transferability property of adversarial examples. Also, an extensive analysis with different values of $\epsilon$, that quantifies the stealthiness of attack, is ignored in [18]. Thus, the impact of adversarial attacks on regression-based PdM models is still under-explored yet. In our work, we study the impact of both white-box and black-box adversarial attacks with different values of $\epsilon$ and also, present a transferability analysis for different DL based regression models in PdM.

Since DL algorithms are popular in many application domains, their vulnerability to cyber attacks is a major concern. In the recent past, many strategies have been proposed [40] for building a robust model against the cyber attacks. For example, adversarial defense through data modification, changes in model structure and using auxiliary tools. The former strategy refers to the reconstruction of training data using adversarial examples, which are crafted using different cyber attacks. It is commonly known as adversarial training [55]. The adversarial examples are added to the training data to make DL model aware of the most expected attacks. Other data modifying techniques include data compression [9], gradient hiding [37], and data randomization [60]. The second strategy refers to enhancing the target neural network for better adversarial robustness. This includes techniques like feature squeezing [62], regularization [7], distillation [38], and mask defense [13]. The third strategy refers to the auxiliary tools such as, MagNet [32], GAN [46], and high-level representation guided denoiser [28]. Among these, GAN is quite popular for defense which generates data similar to accurate data and continues to train itself by producing its own training data. This helps in better modeling of data distribution. However, GAN is hard to train; thus, the network may remain unstable. Therefore, GAN requires good synchronization between the generator and the discriminator. Furthermore, GAN is also prone to *mode collapse* issue, which refers to a generator model that can only generate one or a small subset of different outcomes or modes. Here, *mode* refers to an output distribution, e.g., a multi-modal function refers to a function with more than one peak or optima. With a GAN generator model, a mode failure means that the vast number of points in the input latent space (e.g., hypersphere of 100 dimensions in many cases) result in one or a small subset of generated images. In contrast to GAN, the proposed *approximate adversarial training* technique in this paper for PdM adversarial defense does not employ any discriminators. Therefore, it is easier to train. Also, the network remains stable and free from mode collapse issues.

## 3. Proposed RobustPdM methodology

In this section, a novel methodology is proposed for exploring the impact of adversarial attacks and building a robust RobustPdM system. This methodology can help a PdM designer/developer to analyze not only the accuracy but also the robustness of a PdM system. It is comprised of three key steps, i.e. (1) building a PdM model, (2) crafting adversarial attacks, and, (3) adversarial defense. The detail of these steps is provided in the following section with reference to Fig. 2.

### 3.1. Building a PdM Model

The foremost step of our proposed methodology is to develop a PdM system. A PdM system is a complex system of interconnected components employing IoT sensors and state-of-the-art DL algorithms to make informed maintenance decisions through their efficient training and testing. It uses a repository of prerecorded time-series IoT sensor data, known as *historical dataset*, for training and testing purposes. It is worth mentioning that, in real-world PdM systems, real-time data from the IoT sensors is sequentially recorded as historical data for periodically (or when needed) retraining the PdM model. The main elements of an efficient PdM model designing procedure are described below:

**PdM Model Training** is the most important block in an efficient PdM model building process as the trained PdM model is used for the state prediction of equipment later on. To build an efficient PdM model, the quality of the training data is crucial. The raw sensor readings can be noisy, incomplete, and inconsistent. To make the data ready for building the PdM model, the following pre-processing methods on the sensing data are first adopted:

- *Data cleaning*: Detects and fills missing values in the sensing data and also, detects and removes noisy data points and outliers.
- *Data transformation*: Normalizes IoT sensor data to reduce the data dimensions for its smooth processing.
- *Data reduction*: Samples data records for easier data handling and hence, reduces the computational power required for handling the data.
- *Data discretization*: Convert continuous attributes to categorical attributes for ease of use. It is also known as binning.

Next, the *Analyze and Transform data* block deals with selecting important features in the pre-processed data. Feature selection is the process of identifying the most influential features. Some features may be less important and some may be pivotal for predicting the equipment failure. The key for important features' selection is to understand the role played by every feature, especially when working with a huge amount of data. Reducing the features will result in reducing the computational power needed to train and run the model, which in turn saves a lot of precious time.

Lastly, *Model training* block utilizes a user-defined DL algorithm from the library of DL models and trains it as a PdM model. The PdM designers typically select a DL algorithm based on data and desired accuracy. They use training samples, from the historical dataset, in order to evaluate the performance of the trained model. The evaluation is typically performed using standard metrics such as precision, recall, and F1 and also, error metrics e.g., Mean Absolute error (MAE), Root Mean Square Error (RMSE), and Mean Square Error (MSE). In this paper, we use RMSE which is the most widely used error performance metric in regression models, specifically in PdM applications. For better performance, the model retraining can be scheduled at regular intervals to accommodate the changes in the sensing data with time.

**PdM Model Testing** uses the trained PdM model and testing samples, from the historical dataset, for making the predictions. It first feeds the testing samples to the *data pre-processing* block to handle incomplete and inconsistent testing samples. In the next phase, the prepared testing samples are sent to the *prediction* block, where the trained PdM model makes predictions on them. The prediction provides an RMSE value ($e$), which is helpful in validating the better performance of the PdM system.

*3.2. Crafting adversarial examples*

Once the PdM model is built and tested, the adversarial examples are then crafted for that model to analyze its adversarial robustness. This step takes three inputs: a trained DL model, test data, and library of cyber attacks. A PdM system developer/designer can choose between the state-of-the-art adversarial attacks. In this paper, we adopt four algorithms, MTS FGSM, MTS BIM, MTS PGD and MTS PGD_r from the computer vision domain, convert them for MTS regression models and then apply them to PdM models. The formalization of these adversarial attacks and their corresponding algorithms are explained in detail in Section 4. The adversarial examples that are crafted are given as an input to the PdM model under attack (adversarial PdM model). An adversarial attack on a PdM model is expected to give an RMSE $e'$ which is greater than the RMSE in the previous step (without any attack), i.e. $e' > e$. If the $e'$ is significantly larger than $e$ then it shows that the more PdM model is significantly vulnerable.

*3.3. Adversarial Defense*

As mentioned in Section 2.2.2, there are many strategies in the computer vision domain for adversarial defense against the adversarial examples. In this paper, we propose a novel approximate adversarial training method to design robust PdM systems. In adversarial training, the PdM model is trained on both the training samples from historical data and the previously crafted adversarial examples. This helps the PdM model to learn the adversarial behavior from the the data distribution of the adversarial examples and become more robust. For example, when adversarial training is performed on PdM models with adversarial examples, this results into the RUL prediction with RMSE ($\hat{e}$) closer to the true RUL prediction (without any adversarial defenses), when compared to the RMSE under adversarial attack (without any attack), i.e. $e < \hat{e} < e'$. In approximate adversarial training also, the PdM model learns from the training samples from the historical data and previously crafted adversarial examples. However, unlike non-approximate adversarial training, it incorporates quadratic approximations of the model weights and averages the loss gradient over all batches of adversarial examples in each epoch. Such an approximation and averaging loss gradient concept helps in achieving the robustness higher than non-approximate adversarial training as discussed in Section 4.3.

**4. Adversarial Attacks and Defense in PdM**

In this section, we formalize the problem of adversarial attacks in PdM domain and present the adversarial example generation algorithms.

*4.1. Formalization of the problem*

In a PdM system, an equipment may contain several sensors for different sensor measurements. These measurements are recorded at every time step. Therefore, they can be represented as a MTS data [35].

*Definition 1:* Let $N$ be the number of sensors in an equipment. Then, the MTS sensing data from these sensors can be interpreted as a sequence such that $X = [x_1, x_2, ..., x_T]$, where $T = |X|$ represents the length of $X$ and $x_i \in \mathbb{R}^N$ is a $N$ dimensional data

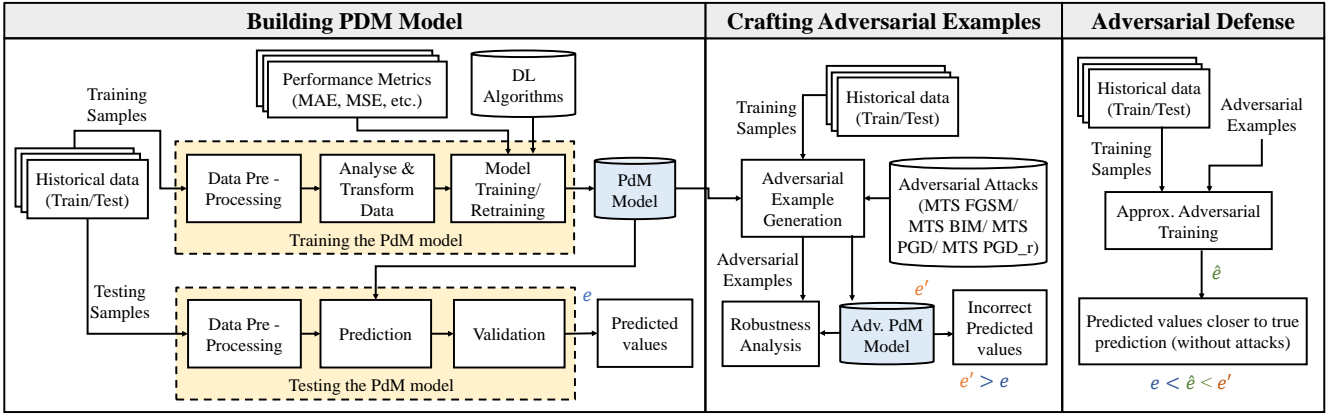

Figure 2: The proposed RobustPdM methodology for designing robust PdM systems

point at time $i \in [1, T]$, which represents the sensor measurements.

*Definition 2:* $D = (x_1, RUL_1), (x_2, RUL_2), ..., (x_T, RUL_T)$ is the dataset of pairs $(x_i, RUL_i)$, where $RUL_i$ is a label (RUL value at that time instant) corresponding to $x_i$.

*Definition 3:* Time series regression task consists of training model with the dataset $D$ in order to predict $R\hat{U}L$ from the possible inputs. $f(\cdot) : \mathbb{R}^{T \times N} \rightarrow R\hat{U}L$ represents a DL model for regression and $J_f(\cdot, \cdot)$ denotes the cost function of model $f$.

*Definition 4:* $X'$ denotes the adversarial example, a perturbed version of $X$ such that $R\hat{U}L \neq R\hat{U}L'$ and $\|X - X'\| \leq \epsilon$. where $\epsilon \geq 0 \in \mathbb{R}$ is a maximum perturbation magnitude.

Given a trained DL model $f$ and an input MTS $X$, crafting an adversarial example $X'$ can be described as the following box-constrained optimization problem.

$$\min_{X'} \|X' - X\| \; s.t.$$

$$f(X') = R\hat{U}L', \; f(X) = R\hat{U}L \text{ and } R\hat{U}L \neq R\hat{U}L' \quad (1)$$

**Adversarial robustness**: Lets assume that a given DL model $f(\cdot)$ has an RMSE $e$ without attack but RMSE $e'$ under attack. Moreover, the model $f(\cdot)$ has an RMSE of $\hat{e}$ after adopting a defense strategy against the adversarial examples, such as non-approximate and approximate adversarial training. Then, the adversarial robustness of $f(\cdot)$ with and without such adversarial defense can be represented as $\beta$ and $\alpha$, respectively, and defined as follows:

$$\alpha = e' - e, \beta = \hat{e} - e \text{ where } \alpha \neq \beta, \text{ and } e < \hat{e} < e' \quad (2)$$

The aim of any adversarial defense should be to find a strategy that minimizes $\beta$.

*4.2. Adversarial example generation*

In this section, we present four adversarial example generation algorithms i.e., MTS FGSM, MTS BIM, MTS PGD and MTS PGD_r attacks. Once again, it is worth mentioning that adversarial attacks from the computer vision domain cannot be directly applied to the PdM domain due to the nature of the attacks (pixels vs. time-series, classification vs. regression). Fundamentally, the regression tasks (which is the case for PdM) use mean squared error (MSE) loss in contrast to the classification problems that use cross-entropy loss to minimize the expected value of the cost function in the training phase. Moreover, in computer vision, small perturbations are added to the images' pixels, leading to the misclassification of the images. In contrast, in the PdM domain, the attackers' objective is to add small perturbations to IoT sensor readings leading to erroneous RUL prediction. Thus, we adopt adversarial example generation algorithms from the computer vision domain for classification tasks and covert them for application to the PdM domain as a multivariate time series regression task. The detail of these proposed attacks are presented as follows.

*Multivariate Time Series Fast Gradient Sign Method (MTS FGSM)*: FGSM was initially proposed to craft the adversarial examples for fooling the GoogleNet model [16]. The FGSM is also named as one shot method as it generates the adversarial examples by a single step computation. MTS FGSM works similar to FGSM by perturbing the multivariate time series input $x$ with one step gradient update along the direction of the sign of gradient at each time step. Mathematically, this can be written as: $\eta = \epsilon \cdot sign(\nabla_x J_f(X, R\hat{U}L))$, where $J_f$ denotes the cost function of model $f$, $\nabla_x$ represents the gradient of the model with respect to the original MTS $X$ with the correct label $R\hat{U}L$, $\epsilon$ is the hyper-parameter which controls the amount of input perturbation and $X'$ is the adversarial MTS. Algorithm 1 delineates the steps of the MTS FGSM attack.

*Multivariate Time Series Basic Iterative Method (MTS BIM)*: The BIM [25] is also known as Iterative-FGSM because it applies FGSM multiple times with small step size. Similar to BIM, MTS BIM clips the original MTS $X$ after each step to en-

| **Algorithm 1:** MTS FGSM algorithm for PdM |
|---|
| **Inputs** : Original multivariate time series $X$ from an equipment and its corresponding label $R\hat{U}L$ |
| **Outputs** : Perturbed multivariate time series $X'$ |
| **Parameter** : $\epsilon$ |
| 1: $\eta = \epsilon \cdot sign(\nabla_x J_f(X, R\hat{U}L))$ |
| 2: $X' = X + \eta$ |

sure that they lie in the range $[X-\epsilon, X+\epsilon]$ i.e. $\epsilon-neighbourhood$ of the original time series $X$. Algorithm 2 presents different steps for the MTS BIM attack. The algorithm requires three hyper-parameters i.e., per step small perturbation ($\alpha$), amount of maximum perturbation ($\epsilon$), and number of iterations ($I$). The value of $\alpha$ is calculated using $\alpha = \epsilon/I$. The adversarial examples generated through this algorithm are usually more closer to the original input and hence, they have a higher chance of fooling the network when compared to MTS FGSM. However, MTS BIM is computationally more expensive and slower than MTS FGSM.

| **Algorithm 2:** MTS BIM algorithm for PdM |
|---|
| **Inputs** : Original multivariate time series $X$ from an equipment and its corresponding label $R\hat{U}L$ |
| **Outputs** : Perturbed multivariate time series $X'$ |
| **Parameter** : $I, \epsilon, \alpha$ |
| 1: $X' \leftarrow X$ |
| 2: **for** i = 1 : I **do** |
| 3:   $\eta = \alpha \cdot sign(\nabla_x J_f(X', R\hat{U}L))$ |
| 4:   $X' = X' + \eta$ |
| 5:   $X' = min\{X + \epsilon, max\{X - \epsilon, X'\}\}$ |
| 6: **end for** |

*Multivariate Time Series Projected Gradient Descent (MTS PGD)*: Unlike MTS BIM, which initializes the adversarial examples generation from the original position, MTS PGD starts from a random point in the $l_\infty$ ball of interest defined by $\delta \leq \epsilon$ where the $\delta$ represents the amount of perturbation. Though this helps MTS PGD achieve a minor improvement over MTS BIM, but overcomes the problem of finding local optima within the objective when MTS PGD starts from a fixed point. In this paper, we crafted the adversarial examples for MTS PGD with two scenarios, (i) simple MTS PGD without restart and (ii) MTS PGD with restart, i.e., MTS PGD_r, to observe their impact on PdM. MTS PGD follows the same Algorithm 3 as MTS PGD_r except that the restart loop is not used in MTS PGD.

### 4.3. Defense through Approximate Adversarial Training

We propose a novel 'approximate adversarial training' as a defense strategy against adversarial attacks to build robust PdM systems. Our technique is inspired by the recent use of approximate computing hardware in deep neural network (DNN) accelerators to offer defense against hardware attack [17]. The authors in [17] designed and utilized an approximate multiplier instead of an accurate multiplier to induce some noise in the DNN accelerator to compensate for the impact of the adversar-

| **Algorithm 3:** MTS PGD_r algorithm for PdM |
|---|
| **Inputs** : Original multivariate time series $X$ from an equipment and its corresponding label $R\hat{U}L$ |
| **Outputs** : Perturbed multivariate time series $X'$ |
| **Parameter** : $I, \epsilon, \alpha$, restarts |
| 1: $X' \leftarrow X$ |
| 2: $J_f^{max} = RUL[0]$ |
| 3: $max\_X' = X$ |
| 4: **for** i = 1 : restarts **do** |
| 5:   $X' = (X * 2 * \epsilon) - \epsilon$ |
| 6:   **for** j = 1 : I **do** |
| 7:     $X' = min\{X + \epsilon, max\{X - \epsilon, X'\}\}$ |
| 8:   **end for** |
| 9:   $all\_loss = \nabla_x J_f(X, RUL)$ |
| 10:  $max\_X' = X[all\_loss \geq J_f^{max}]$ |
| 11:  $J_f^{max} = max(J_f^{max}, all\_loss)$ |
| 12: **end for** |

ial attacks. Our proposed method uses a software-based approximation technique compared to a hardware-based approximation. Specifically, our proposed approximate adversarial training comprises three key steps, i.e., adversarial examples generation, averaging the loss gradient, and weight approximation. As shown in Algorithm 4, we first generate a set of adversarial examples $x^*$ and a set of their corresponding labels $y^*$ under all attacks on a model $\theta$ by appending them in an empty list next to each other (Line 1-8). Then, we train the PdM model against these adversarial examples by dividing these two sets into $k$ number of batches $b$ (Line 11). In each training epoch, we average the loss gradient $\nabla_x J_f$ over all batches of adversarial examples (Line 13), approximate the weights by dividing them into uniform $m$ groups and apply the quadratic approximation methods to each weights group $w_g$. Specifically, we compute the quadratic coefficients $q0$, $q1$ and $q_2$ by applying the quadratic function i.e., $q_0 + xq_1 + x^2q_2$, where $x$ represents the initial weights. Using the coefficients to calculate the new approximate weights $w_g^a$ (Line 14). After this, we update the model using the averaged loss gradient in each training epoch (Line 15). Lastly, we return the updated model $\theta'$ (Line 18) to evaluate the performance of non-approximate adversarial training in the prediction. Note, the classical 'adversarial training' [55] (termed as non-approximate adversarial training in this paper) lacks averaging the loss gradient over all adversarial examples and does not employ weight approximation. It is worth mentioning that the weight approximation improves the robustness of PdM models by introducing approximation errors in the model. An adversary may not be aware of such approximation in the model; hence, the attacks on a model trained with approximate adversarial examples do not significantly impact.

### 5. Case study: Turbofan Engine PdM

In this section, we first build PdM models with four different DL algorithms using the proposed methodology. This step is required since without PdM models, it is not possible to measure the impact of adversarial examples on such models. Then,

Table 1: Architecture and hyper-parameters for different DL algorithms

| DL architecture | Hidden neurons | Batch Size | Epochs | RMSE |
|---|---|---|---|---|
| CNN(64,64,64,64) lh(100) | 64,64,64,64 | 256 | 120 | 9.93 |
| LSTM(100,100,100,100) lh(80) | 100,100,100,100 | 200 | 100 | 8.80 |
| GRU(100,100,100) lh(80) | 100,100,100 | 200 | 150 | 7.62 |
| Bi-LSTM(180,180,120) lh(60) | 180,180,120 | 200 | 100 | 8.43 |

**Algorithm 4:** Approximate Adversarial Training

**Inputs** : Model $\theta$, list of attacks *attacks*, Multivariate time series $\mathcal{D} = (x, y)$
**Outputs** : Adversarially trained model $\theta'$
**Parameter** : $\epsilon$, *epochs*, batches $n$, batch size $b$, learning rate $\eta$

1: $x^* = [\ ]$
2: $y^* = [\ ]$
3: **for** $i = 1 : size(x^*)$ **do**
4:    **for** $j = 1 : size(\mathcal{D})$ **do**
5:       $(xa_j^*, ya_j^*) = \text{GenAdvExamples}(attacks[i], \epsilon, \theta, x_j, y_j)$
6:       $x_j^* = [x_j^*, xa_j^*]$
7:       $y_j^* = [y_j^*, ya_j^*]$
8:    **end for**
9: **end for**
10: **for** $k = 1 : n$ **do**
11:    $b_k^* = \{(x_l^*, y_l^*) : l \in [1, m]\}$
12:    **for** $e = 1 : epochs$ **do**
13:       $\mathcal{L}_k = (1/b) \sum_{j=1}^{b} \nabla_x J_f(b_k^*, \theta)$
14:       $w_g^a = \text{QuadraticApprox}(w_{g_i}), \forall i \in [1, m]$
15:       $\theta = \theta + \eta \mathcal{L}_k w_g^a$
16:    **end for**
17: **end for**
18: $\theta' = \theta$
19: **return** $\theta'$

following the methodology we craft adversarial examples for the developed PdM models and evaluate their impact on those models. Finally, we perform adversarial training to enhance the adversarial robustness of the PdM models and analyze them.

### 5.1. Building PdM models for turbofan engine

For building the PdM models, we use NASA's turbofan Commercial Modular Aero-Propulsion System Simulation dataset, known as C-MAPSS [42]. Specifically, we use the FD001 sub-dataset from C-MAPSS for our experiments which include 21 sensors data with a different number of operating conditions and fault conditions. FD001 contains training and testing data for 100 engines. The test data contains run to failure data from several engines of the same type. Each row in test data is a time cycle which can be defined as an hour of operation. A time cycle has 26 columns where the first column represents the engine ID, and the second column represents the current operational cycle number. The columns from 3 to 5 represent three operational settings. The columns from 6 to 26 represent the 21 sensor values. The time-series data terminates only when a fault is encountered. For example, an engine with ID 1 has 192 time cycles of data, which means the engine has developed a fault at the 192 time cycle. The test data contains data only for some time cycles as our goal is to estimate the remaining operational time cycles before a fault. Note, 7 out of these 21 engines can be ignored since their measurements remain constant. For the rest of the 14 sensors, we used the normalization technique to convert the raw sensory data into a normalized scale. Figure 3 shows a 3-D representation of the sensor data from engine ID 49 for 300 time cycles. In this paper, we used the resultant normalized dataset to generate adversarial examples using MTS FGSM, MTS BIM, MTS PGD and MTS PGD_r.

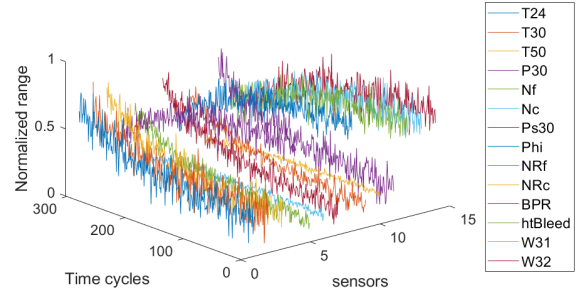

Figure 3: Normalized sensor measurements

We use four DL models specifically, CNN, LSTM, Bi-LSTM, and GRU for predicting the RUL of the aircraft engines as they are known for their applicability in the PdM domain [49, 43]. The architecture, hyperparameters and RMSE of these models are shown in Table 1. It shows that when there is no attack present, GRU (100, 100, 100) with a sequence length 80 provides the most accurate RUL prediction with the least RMSE of 7.62 among these four evaluated models. The notation GRU (100, 100, 100) lh(80) refers to a network that has 100 nodes in the hidden layers of the first, second, and third GRU layers, and a sequence length of 80. Next, we craft adversarial examples for these models and evaluate their impacts by comparing the RMSE of these DL models.

### 5.2. Threat model for the turbofan engine PdM

Before proceeding to the results, we describe the threat model for the turbofan engine PdM case study as follows.

**Attack objective:** The objective of the attacker is to trigger either an early or delayed maintenance. An early or, in other words, unnecessary maintenance can result into flight downtime and unnecessary maintenance. Both of them can lead to a loss of flight time, loss of human effort, loss of resources, and also incur an extra maintenance cost. On the other hand, delayed maintenance can lead to an engine failure and also, the loss of human lives in the worst case.

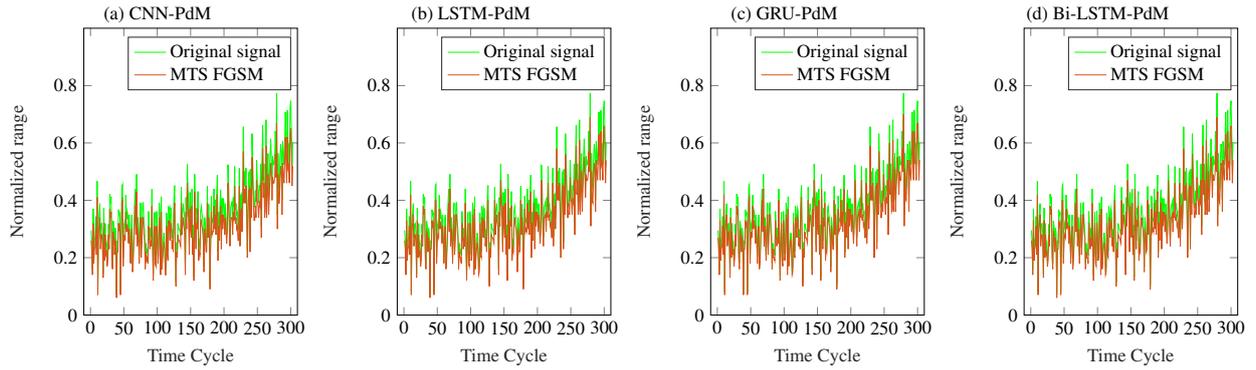

Figure 4: MTS FGSM ($\epsilon = 0.3$) attack signatures for sensor 2 of engine ID 49 for DL-based PdMs

**Attack surface:** In this work, we consider both, the white-box and black-box attacks. In both cases, we use only the crafted adversarial examples and we do not perturb the model parameters at all. In the white-box attack, the adversary has access to the data and has the knowledge of internal parameters of the DL model. In contrast, in black-box attack, the adversary has access to the data but has no knowledge of employed DL models. The results of the black-box attack are demonstrated through the transferability property of the adversarial examples.

In the case of aircraft engine's predictive maintenance, the attacker can access the sensor data by exploiting the controlled area network (CAN) bus systems aboard aircraft and feed the crafted adversarial examples through a false data injection (FDI) [41] attack. The FDIA attack can be launched through spoofing techniques. Tippenhauer et al. presented a spoof attack scenario on GPS-enabled devices [54]. In this attack scenario, a forged GPS signal is transmitted to the device to alter the location. In this way, the true location of the device is disguised and the attacker can perform a physical attack on the device. In another work, Giannetos et al. introduced an app named 'Spy-sense', which monitors the behaviors of several sensors in a device [14]. The app can manipulate sensor data by deleting or modifying it. Spy-sense exploits the active memory region in a device and relays sensitive data covertly. These works demonstrates that FDIA attacks can be performed even without gaining direct access to a system.

One of the recent articles [26] considers cyber-attacks as one of the potential reasons behind the two recent Boeing 737 Max 8 crashes. According to this article, a passenger or vehicle or drone carrying a sonic device capable of impacting the MCAS sensor controlling the plane could have been responsible for such an attack. Recently, ICS-CERT published an alert on certain controlled area network (CAN) bus systems aboard aircraft that may be vulnerable to hacking. It cited a report that an attacker with access to the aircraft could attach a device to avionics CAN bus to inject false data, resulting into incorrect readings in an avionic equipment [57]. Using such a device attached to the bus could lead to incorrect engine telemetry readings, compass, altitude, airspeed, and angle of attack (AoA) data. Pilots or maintenance engineer, on the ground, may not be able to distinguish between false and legitimate readings since the (adversarial) attack signature can be very stealthy in nature (as discussed in Section 5.3). This alert explores the possibility of injecting adversarial examples as false data into IoT sensor readings of aircraft engine which are transmitted on a CAN. In this work, we consider FDIA using a malicious device attached to avionics CAN.

*5.3. Crafting stealthy adversarial examples*

To analyze the impact of adversarial attacks on turbofan engine PdM, we create a subset of test data from FD001 in which each engine has at least 150 time cycles of data. This gives us 37 engines in the FD001 dataset. This is done because more time cycles of data in an engine helps the DL models to make more accurate RUL predictions. The resultant dataset is re-evaluated using the Bi-LSTM, LSTM, CNN, and GRU-based PdM models and the obtained RMSEs are 5.81, 5.83, 7.92, and 5.77, respectively. To analyze the impact of white-box adversarial attacks on the C-MAPSS, we craft the adversarial examples for MTS FGSM, MTS BIM, MTS PGD and MTS PGD_r attacks, as discussed in Section 4.2, and apply them on the DL models. For illustration, Fig. 4, Fig. 5, Fig. 6 and Fig. 7 show the examples of a perturbed data from sensor 2 of engine ID 49 crafted using MTS FGSM (with $\epsilon = 0.3$), MTS BIM (with $\alpha = 0.003$, PGD and PGD_r, $\epsilon = 0.3$, $I = 100$ and restarts = 30), respectively. We choose $\epsilon = 0.3$ for MTS FGSM, MTS BIM, MTS PGD and MTS PGD_r attacks to make sure that the crafted adversarial examples are stealthy. Thus, as shown in Figs. 4–7, the crafted/peturbed signals look very similar to the original input signals. Such stealthy attacks often fall within the boundary conditions of the sensor measurements and hence, they are indeed hard to detect using the common attack detection mechanisms.

From Fig. 8, we observe that the MTS FGSM attack (with $\epsilon = 0.3$) increases the RMSE of CNN, LSTM, GRU, and Bi-LSTM models by 2.3X, 2.34X, 1.94X, and 2.19X, respectively, when compared to the PdM models without attack. For the MTS BIM attack (with $\alpha = 0.003$, $\epsilon = 0.3$ and $I = 100$), we observe a similar trend i.e., the RMSE for the CNN, LSTM, GRU, and Bi-LSTM model is increased by 3.94X, 4.51X, 4.46X, and 4.51X, respectively, when compared to the PdM models without attack. Furthermore, MTS PGD and MTS PGD_r increase

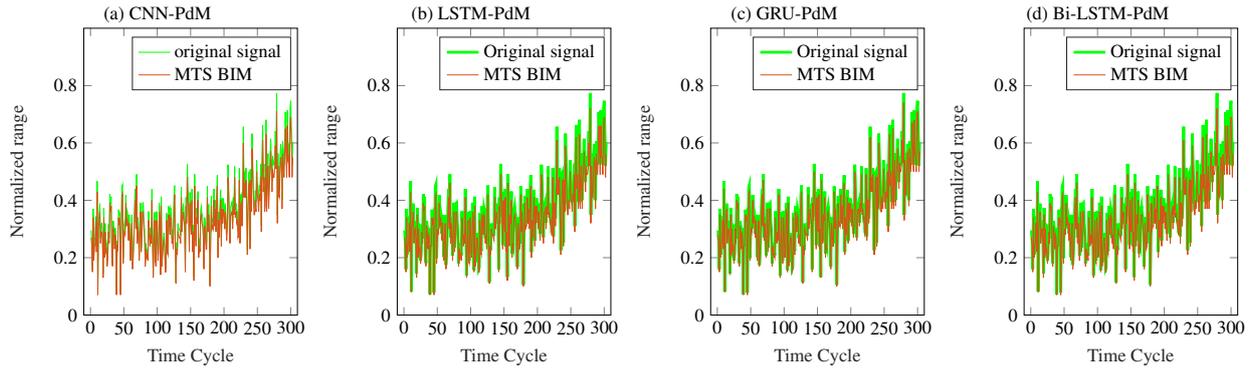

Figure 5: MTS BIM ($\alpha = 0.003$, $\epsilon = 0.3$, and $I = 100$) attack signature for sensor 2 of engine ID 49 for DL-based PdMs

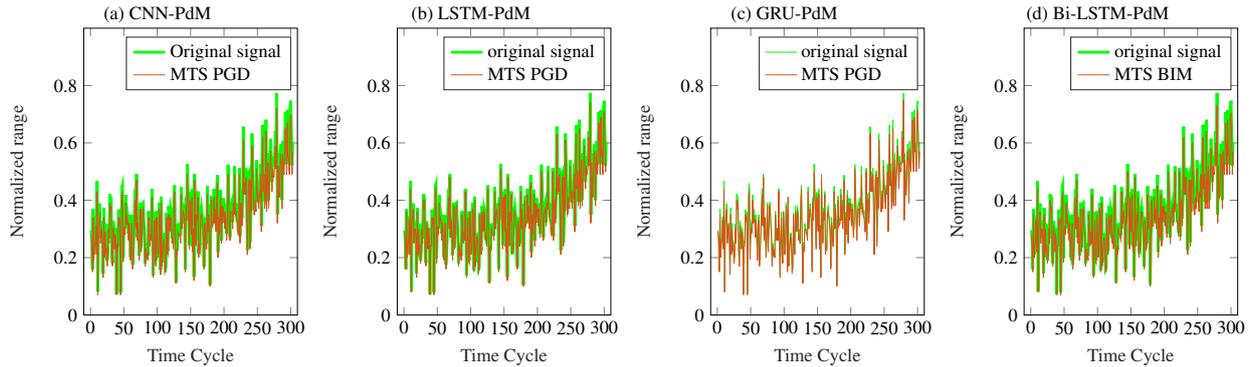

Figure 6: MTS PGD ($\alpha = 0.003$, $\epsilon = 0.3$, and $I = 100$) attack signature for sensor 2 of engine ID 49 for DL-based PdMs

RMSE to a larger extent. For example, 5.16X, 6.35X, 6.26X, and 6.38X increases in RMSE for MTS PGD, and 5.72X, 7.26X, 7X, and 7.06X increases in RMSE for MTS PGD_r for the CNN, LSTM, GRU, and Bi-LSTM is observed, compared to the PdM models without attack.

### 5.4. Impact of attacks on RUL estimation

As mentioned in the attacker's objective, MTS FGSM, MTS BIM, MTS PGD, and MTS PGD_r attacks can cause an under-prediction or over-prediction. For instance, as shown in Fig. 8, the CNN model predicts the RUL (without attack) of 125 (in hours) for engine ID 33 and 132 (in hours) for engine ID 4. After performing MTS FGSM, MTS BIM, MTS PGD, and MTS PGD_r attacks for engine ID 33, the same CNN model predicts the RUL (in hours) as 176, 250, 234, and 239, respectively. This represents a 1.4X, 2X, 1.89X, and 1.92X increase in RUL after MTS FGSM, MTS BIM, MTS PGD, and MTS PGD_r attacks. For engine ID 4, the MTS FGSM, MTS BIM MTS PGD, and MTS PGD_r attacks result in RUL of 97, 78, 70, and 75, respectively. This represents a 1.26X, 1.69X, 1.88X, and 1.77X decrease in the predicted RUL after MTS FGSM, MTS BIM, MTS PGD, and MTS PGD_r attacks. An over-prediction, as shown in engine ID 33, may cause delayed maintenance, whereas an under-prediction, as shown in engine ID 4, may cause early maintenance. Both of them have catastrophic consequences.

To elucidate the impact of MTS FGSM, MTS BIM, MTS PGD and MTS PGD_r attacks on specific engine data, we first apply the piece-wise RUL prediction (using the same DL models) for a single-engine (engine ID 17 in this case) and then, apply the crafted adversarial examples. The piece-wise RUL prediction gives a better visual representation of degradation in an aircraft engine. Fig. 9 shows the results for the piece-wise RUL prediction using CNN, LSTM, GRU, and Bi-LSTM models. It is evident from this figure that when time approaches the end of life, the predicted RUL (green solid line) gets closer to the true RUL (black dashes). This is because, the RUL predictions get more accurate with the increasing amount of data.

Next, we craft adversarial examples using MTS FGSM, MTS BIM, MTS PGD and MTS PGD_r for the engine ID 17, apply them for piece-wise RUL prediction and compare their impact (as shown in Fig. 9). We observe that the crafted adversarial examples have a strong impact from the beginning of the RUL prediction on the CNN model when compared to the Bi-LSTM, LSTM and GRU models. The piece-wise RUL prediction after the attack on the CNN model follows the same trend of the piece-wise RUL without attack. However, the attacked RUL values remain quite far from the actual prediction. On the other hand, the impact of adversarial attacks on the GRU model is quite interesting. We observe that the RUL remains almost constant up to 104, 129, 140 and 145 time cycles for MTS FGSM, MTS BIM, MTS PGD and MTS PGD_r attacks, respectively, then starts decreasing. Such a phenomenon is deceiving in nature. It indicates that the engine is quite healthy and may influence a 'no maintenance required' decision by the maintenance engineer. Once again, it is evident that the PGD_r attack has a stronger impact on piece-wise RUL prediction when compared

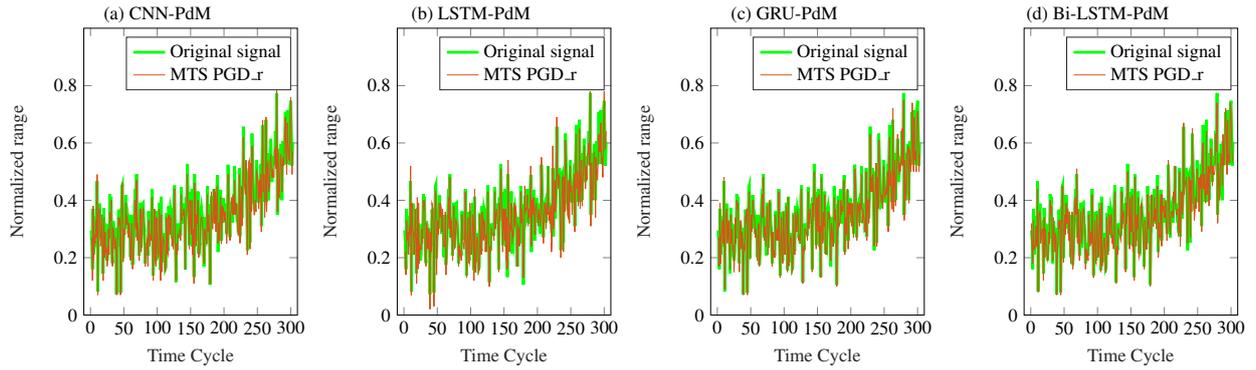

Figure 7: MTS PGD_r ($\alpha = 0.003$, $\epsilon = 0.3$, $I = 100$, *restart*=20) attack signature for sensor 2 of engine ID 49 in PdMs

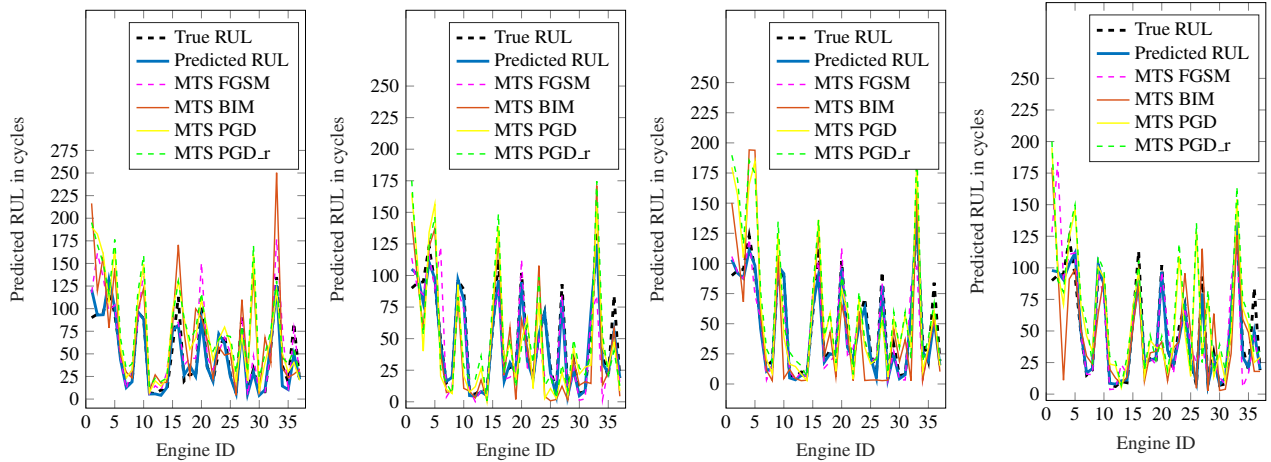

(a) CNN-PdM under MTS FGSM (RMSE = 18.35), MTS BIM (RMSE = 31.23), MTS PGD (RMSE = 40.94) and MTS PGD_r (RMSE = 45.38) attack

(b) LSTM-PdM under MTS FGSM (RMSE = 13.65), MTS BIM (RMSE = 26.35) and MTS PGD (RMSE = 37.07) and MTS PGD_r (RMSE = 42.38) attack

(c) GRU-PdM under MTS FGSM (RMSE = 11.22), MTS BIM (RMSE = 25.79) and MTS PGD (RMSE = 36.14) and MTS PGD_r (RMSE = 40.56) attack

(d) Bi-LSTM-PdM under MTS FGSM (RMSE = 12.77), MTS BIM (RMSE = 26.23) and MTS PGD (RMSE = 36.97) and MTS PGD_r (RMSE = 41.03) attack

Figure 8: RUL estimation under MTS FGSM ($\epsilon = 0.3$), MTS BIM ($\alpha = 0.003$, $\epsilon = 0.3$, and $I = 100$), MTS PGD ($\alpha = 0.003$, $\epsilon = 0.3$, and $I = 100$) and MTS PGD_r ($\alpha = 0.003$, $\epsilon = 0.3$, and $I = 100$, $restart = 30$) attacks

to the MTS FGSM, MTS BIM and MTS PGD attacks.

**Transferability of attacks:** We perform a comprehensive transferability analysis for evaluating the impact of black-box attacks on PdM models. Specifically, we apply the adversarial examples crafted for a PdM model to the other PdM models using the data of 37 engines, as mentioned in Section 5.1. In black box attacks, the attacker has no knowledge about the target model's internal parameters [37], but it can still have a considerable impact on the target model. Our obtained results are presented in Table 2. The first column (DL models) of the Table 2 represents the RMSE of the models without attack. We observe that the MTS FGSM, MTS BIM, MTS PGD, and MTS PGD_r adversarial examples crafted for the CNN model give a higher RMSE when transferred to other DL models. Interestingly, when the adversarial examples are crafted using MTS PGD_r, they show better transferability by inducing a higher RMSE. For instance, the CNN-based PdM model has an RMSE of 7.92. When we craft adversarial examples for the CNN model using MTS FGSM, MTS BIM, MTS PGD, and MTS PGD_r and transfer them to the GRU model, we observe that MTS PGD_r adversarial example increases the RMSE (40.11) almost 2.8X, 1.5X, and 1.1X when compared to MTS FGSM (14.23), MTS BIM (27.12), MTS PGD (36.76), respectively. When adversarial attacks are transferred, a similar trend is also observed for Bi-LSTM and other PdM models.

**Performance variation vs. the amount of perturbation:** In these experiments, we explore the impact of the amount of perturbation $\epsilon$ on the GRU model performance in terms of RMSE. We picked the GRU model because it showed the best performance in predicting the RUL. The obtained result is shown in Figure 10. We observe that for larger values of $\epsilon$, MTS PGD and MTS PGD_r attacks result in higher RMSE when compared to the MTS BIM and MTS FGSM. For instance, for $\epsilon = 1.3$, MTS FGSM and MTS BIM attacks result in an RMSE of 32.78 and 61.34, respectively, whereas MTS PGD and MTS PGD_r attacks result in an RMSE of 75.53 and 82.89, respectively. This shows that MTS PGD and MTS PGD_r generates adversarial examples impacting more than 2X when compared to the MTS BIM and MTS FGSM for the same value of $\epsilon$. This is due to the fact that MTS PGD and PGD_r adds a small amount of perturbation $\alpha$ on each iteration randomly whereas MTS FGSM adds the total amount of perturbation $\epsilon$ on each data point [25].

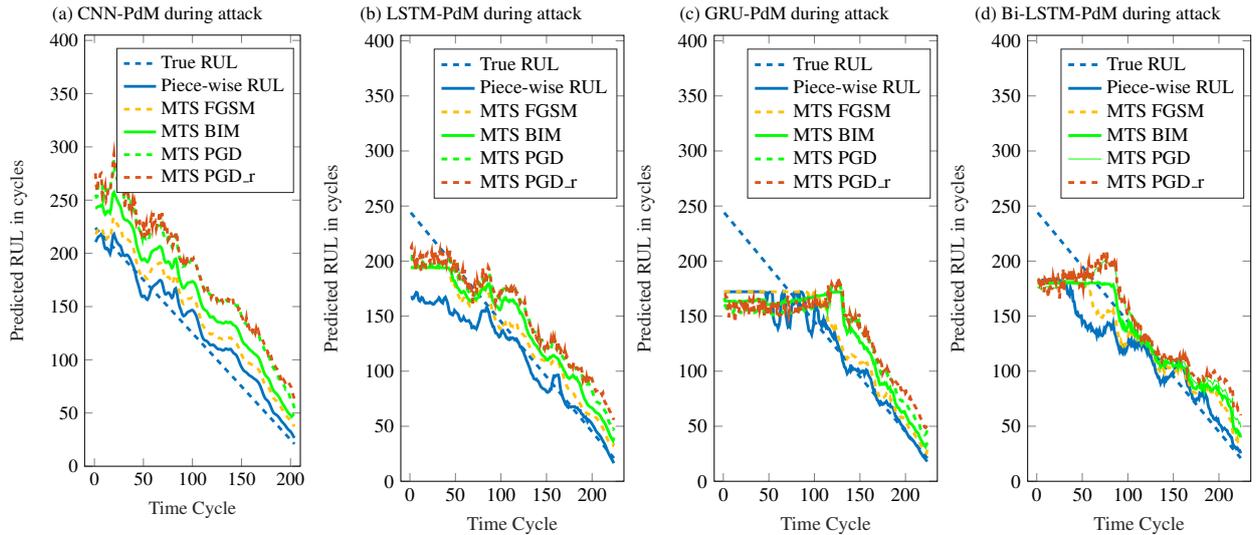

Figure 9: Piece-wise RUL prediction under MTS FGSM ($\epsilon = 0.3$), MTS BIM ($\alpha = 0.003$, $\epsilon = 0.3$, and $I = 100$), MTS PGD ($\alpha = 0.003$, $\epsilon = 0.3$, and $I = 100$) MTS PGD_r attacks ($\alpha = 0.003$, $\epsilon = 0.3$, and $I = 100$ and $restart=100$)

Table 2: Transferability of MTS FGSM, MTS BIM, MTS PGD and MTS PGD_r attacks. The notation X/Y/Z/W represents RMSE using MTS FGSM/MTS BIM/MTS PGD/MTS PGD_r attacks

| DL Models | RMSE | | | |
|---|---|---|---|---|
| | CNN | LSTM | Bi-LSTM | GRU |
| CNN (RMSE = 7.92) | - | 18.76 / 29.45 / 38.65 / 40.45 | 14.44 / 24.62 / 32.54 / 35.43 | 14.23 / 27.12 / 36.76 / 40.11 |
| LSTM (RMSE = 5.83) | 18.23 / 31.13 / 40.57 / 44.75 | - | 14.74 / 21.26 / 37.05 / 39.87 | 11.33 / 18.65 / 27.32 / 30.63 |
| Bi-LSTM (RMSE = 5.81) | 17.44 / 30.67 / 41.53 / 45.66 | 12.32 / 21.36 / 29.43 / 32.65 | - | 10.72 / 19.66 / 25.67 / 28.35 |
| GRU (RMSE = 5.77) | 16.22 / 30.45 / 39.85 / 43.83 | 10.89 / 19.52 / 26.54 / 29.52 | 9.45 / 17.66 / 23.87 / 25.96 | - |

### 5.5. Defense: Approximate adversarial training

From the previous experiments, we observe that adversarial attacks can significantly impact RUL prediction. In this section, we apply the proposed approximate adversarial training in Algorithm 4. Also, we compare it with the traditional non-approximate adversarial training, which is popular in the computer vision domain. We train the PdM models using the adversarial examples crafted for epsilon values in the range of 0.1 to 0.9 and learning rate $\eta = 0.001$. Next, we test these nine adversarially trained models with adversarial examples crafted using $\epsilon = 0.3$. We choose $\epsilon = 0.3$ for crafting adversarial examples because such adversarial examples are stealthy and result in a significant impact on PdM systems, as reported in the previous section. It is worth mentioning that it is a normal phenomenon to have a slight increase in the RMSE when adversarial training is applied for defense [56]. We also observe this when we apply both non-approximate and approximate adversarial training for defense, as shown in Table 3. However, our proposed approximate adversarial training results in a lower percentage increase in RMSE than conventional non-approximate adversarial training.

**Impact of non-approximate adversarial training:** In Table 4, the first column represents four PdM models with their respective values of alpha ($\alpha$) after MTS FGSM, MTS BIM, MTS PGD and MTS PGD_r attacks for $\epsilon = 0.3$. Let us recall the definition of $\alpha$ and $\beta$ from Section 3. The $\alpha$ and $\beta$ are the evaluation metrics for measuring the adversarial robustness before and after adversarial training. For instance, the value of $\alpha$

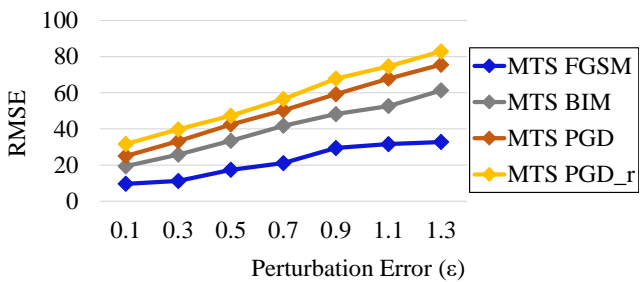

Figure 10: RMSE variation with respect to the amount of perturbation ($\epsilon$) for MTS FGSM, MTS BIM, MTS PGD and MTS PGD_r attacks

Table 3: RMSE comparison before and after adversarial training

| DL architecture | RMSE before adv. training | RMSE after adv. training (non-approx., approx.) | Percentage increase (non-approx., approx.) |
|---|---|---|---|
| CNN | 7.92 | (8.20, 8.08) | (3.53, 2.02) % |
| LSTM | 5.83 | (6.08, 6.03) | (4.28, 3.43)% |
| GRU | 5.77 | (5.97, 5.9) | (3.46, 2.25) % |
| Bi-LSTM | 5.81 | (6.06, 5.95) | (4.30, 2.40)% |

Table 4: Non-approximate adversarial training using MTS FGSM, MTS BIM, MTS PGD and MTS PGD_r attacks. The notation X/Y/Z/W represents $\beta$ of test data after non-approximate adversarial training using MTS FGSM/MTS BIM/MTS PGD/MTS PGD_r attacks

| PdM Models | $\beta$ | | | |
|---|---|---|---|---|
| | $\epsilon = 0.1$ | $\epsilon = 0.3$ | $\epsilon = 0.5$ | $\epsilon = 0.7$ |
| CNN ($\alpha$ : 10.54/23.31/31.45/33.61) | 0.9/1.31/1.63/1.70 | 0.4/0.54/0.70/0.76 | 0.56/0.73/0.89/0.97 | 0.83/1.19/1.35/1.46 |
| LSTM ($\alpha$ : 7.82/20.52/28.45/31.03) | 1.1/1.59/1.84/1.93 | 0.72/0.94/1.09/1.18 | 0.89/1.06/1.20/1.26 | 1.13/1.36/1.54/1.63 |
| Bi-LSTM ($\alpha$ : 6.96/20.42/29.65/32.45) | 1.01/1.56/1.87/1.96 | 0.69/0.91/1.08/1.19 | 0.85/1.0/1.09/1.16 | 1.12/1.3/1.42/1.47 |
| GRU ($\alpha$ : 5.45/20.02/31.78/34.53) | 0.98/1.41/1.73/1.79 | 0.69/0.87/1.19/1.26 | 0.75/1.01/1.26/1.33 | 1.08/1.28/1.42/1.49 |

Table 5: Approximate Adversarial Training using MTS FGSM, MTS BIM, MTS PGD and MTS PGD_r attacks. The notation X/Y/Z/W represents $\beta$ of test data after approximate adversarial training using MTS FGSM/MTS BIM/MTS PGD/MTS PGD_r attacks

| PdM Models | $\beta$ | | | |
|---|---|---|---|---|
| | $\epsilon = 0.1$ | $\epsilon = 0.3$ | $\epsilon = 0.5$ | $\epsilon = 0.7$ |
| CNN ($\alpha$ : 8.95/20.12/29.54/30.98) | 0.79/1.14/1.45/1.49 | 0.29/0.37/0.58/0.59 | 0.45/0.60/0.75/0.85 | 0.74/1.05/1.18/1.29 |
| LSTM ($\alpha$ : 6.98/17.85/25.93/29.80) | 0.98/1.38/1.66/1.71 | 0.61/0.79/0.97/1.08 | 0.80/0.95/1.06/1.11 | 1.05/1.20/1.38/1.49 |
| Bi-LSTM ($\alpha$ : 5.99/19.16/27.15/30.03) | 0.94/1.41/1.72/1.79 | 0.61/0.79/0.95/1.03 | 0.70/0.89/0.93/1.05 | 0.98/1.11/1.27/1.33 |
| GRU ($\alpha$ : 4.91/18.32/29.85/31.06) | 0.87/1.26/1.59/1.64 | 0.58/0.79/1.06/1.14 | 0.64/0.93/1.13/1.19 | 0.94/1.10/1.30/1.40 |

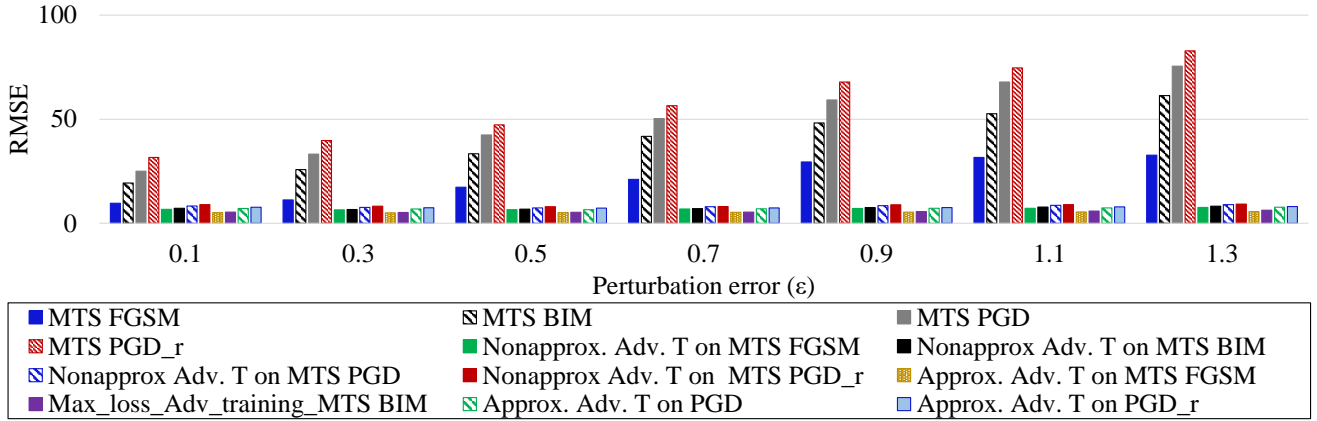

Figure 11: Comparison of adversarial trained models with non-adversarial trained models under adversarial attacks w.r.t. the amount of perturbation ($\epsilon$) for GRU-PdM model

for the CNN model, in the case of an MTS FGSM attack, is calculated as a difference between $e' = 18.32$ and $e = 7.92$, which is equal to 10.40. The value of $\beta$ for the CNN model (trained on MTS FGSM adversarial examples of $\epsilon = 0.3$ and tested on adversarial examples using the same epsilon) is calculated as a difference between $\hat{e} = 8.32$ and $e = 7.92$, which gives us the difference as 0.4. A robust PdM should have smaller values of $\beta$ compared to $\alpha$. From Table 4, it is observed that the effect of an adversarial attack is greatly reduced (thus, significantly improved adversarial robustness) after adversarial training for MTS FGSM, MTS BIM, MTS PGD, and MTS PGD_r attacks, i.e., $\beta < \alpha$ for all PdM models. For example, the values of ($\alpha$, $\beta$) are (10.54, 0.4), (23.31, 0.54), (31.45, 0.7) and (33.61, 0.76) when $\epsilon = 0.3$ for MTS FGSM, MTS BIM, MTS PGD and MTS

PGD_r attacks, respectively, in case of CNN-based PdM model. This shows almost 26X, 43X, 45X, and 44X improvement in robustness under MTS FGSM, MTS BIM, MTS PGD, and MTS PGD_r attacks, respectively. In all PdM models, the values of $\alpha$ are 5.45 to 34.53, whereas the values of the $\beta$ lie from 0.4 to 1.96. Furthermore, we also investigate the performance of non-approximate adversarial training with different values of $\epsilon$. For this analysis, we picked the GRU model because it showed the best performance in predicting the RUL (as discussed in Section 5.3). As shown in Fig. 11, we observe the value of RMSE after MTS FGSM, MTS BIM, MTS PGD, and MTS PGD_r attacks on the GRU model increase with an increase in the value of $\epsilon$. However, the value of RMSE decreases significantly after non-approximate adversarial training. For instance, MTS FGSM, MTS BIM, MTS PGD, and MTS PGD_r attacks, with $\epsilon = 1.3$, on non-adversarial trained model results in the RMSE of 32.78, 61.34, 75.53 and 82.89 respectively. In contrast, the same PdM model shows an RMSE of 7.67, 8.19, 8.95, and 9.21 for MTS FGSM, MTS BIM, MTS PGD, and MTS PGD_r, respectively, after non-approx adversarial training. From these results, we discern that non-approximate adversarial training can successfully improvise the robustness of PdM models.

**Impact of approximate adversarial training:** Similar to non-approximate adversarial training, the effect of an adversarial attack is significantly reduced after approximate adversarial training for MTS FGSM, MTS BIM, MTS PGD, and MTS PGD_r attacks. As shown in Table 5, $\beta < \alpha$ for all PdM models. For example, the values of $(\alpha, \beta)$ are (8.95, 0.29), (20.12, 0.37), (29.54, 0.58) and (30.98, 0.59) when $\epsilon = 0.3$, for MTS FGSM, MTS BIM, MTS PGD and MTS PGD_r attacks, respectively, in case of CNN-based PdM model. This shows almost 31X, 54X, 51X, and 52X improvement in robustness under MTS FGSM, MTS BIM, MTS PGD, and MTS PGD_r attacks, respectively. In all PdM models, the values of $\alpha$ are in the range of 4.91 to 31.06, whereas the values of the $\beta$ lie in the range from 0.29 to 1.79 only. Furthermore, in comparison to non-approximate adversarial training, the values for $\beta$ are much reduced in approximate adversarial training. The difference in adversarial robustness with approximate adversarial training is 11X higher than that with non-approximate adversarial training. This shows significant improvement in PdM robustness with non-approximate adversarial training compared to its non-approximate counterpart. We also investigate the performance of approximate adversarial training with different values of $\epsilon$. As shown in Fig. 11, approximate adversarial training reduces the value of RMSE even with an increase in the values of $\epsilon$ for MTS FGSM, MTS BIM, MTS PGD, and MTS PGD_r attacks on GRU model. Interestingly, this reduction in RMSE with approximate adversarial training is much lower than that with non-approximate adversarial training. For instance, approximate adversarial training of the GRU model results in RMSE of 5.65, 6.29, 7.7, and 8.05 only under MTS FGSM, MTS BIM, MTS PGD, and MTS PGD_r, respectively.This is indeed 5.8X, 9.75X, 9.8X and 10.3X lower than the RMSE in non-approximate adversarial training for MTS FGSM, MTS BIM, MTS PGD, and MTS PGD_r attacks, respectively. This shows significant improvement in PdM robustness achieved via approximate adversarial training.

### 5.6. Discussion and Limitations

To summarize, our results show that the proposed attacks and defense outperform the state-of-the-art attacks and defense methods in PdM. For instance, our proposed MTS PGD and MTS PGD_r attacks result in up to 4X higher RMSE than the state-of-the-art FGSM and BIM attacks in [33]. Also, they result in up to 3X higher RMSE when compared to the state-of-the-art ROM and MIM attacks in [18]. This means that our proposed approximate adversarial defense can be applied to defend PdM models against not only MTS PGD and MTS PGD_r attacks, but also, ROM and MIM attacks. Our proposed approximate adversarial training method improves the robustness of PdM models up to 54X when compared to the PdM models without defense, which is upto 3X more effective than the state-of-the-art defense proposed in [18]. Finally, we also observe that our proposed approximate adversarial training improves the robustness of PdM models by 11X compared to non-approximate adversarial training.

The attack surface of a PdM system is vast. An adversary can target a physical, edge, cloud, or visualization layer in a PdM system. However, our proposed defense is limited to providing a defense against attacks on DL models only, e.g., with adversarial perturbations in IoT sensor data. Therefore, our proposed defense is a part of a broader cyber-security solution toward more robust PdM systems. It is worth mentioning that the vulnerabilities of PdM models are evolving every day. In such a race to counterfeit the impact of every new defense mechanism with stronger attacks, it is imperative to analyze the proposed defense mechanism extensively. Therefore, we plan to further extend our work to a more extensive evaluation in the future by generalizing the proposed approximate adversarial training with new and stronger attacks.

## 6. Conclusion

This paper proposes two new adversarial attacks, a novel defense method, and an end-to-end RobustPdM methodology for building robust PdM systems. Specifically, we proposed MTS PGD and MTS PGD_r attack for the PdM domain and evaluated their impacts in addition to the MTS FGSM and MTS BIM attacks on LSTM, GRU, CNN, and Bi-LSTM based PdM models. Furthermore, we proposed a novel defense technique for approximate adversarial training that applies quadratic coefficients by approximating model weights and averages the loss gradient over multiple batches of adversarial examples. The proposed attack and defense methods were evaluated via the RobustPdM methodology using NASA's turbofan engine case study. The results show that the proposed attacks are 2X more effective than the state-of-the-art PdM attacks. Consequently, we observed that approximate adversarial training successfully defended adversarial attacks in PdM systems. Specifically, approximate adversarial training can improve the robustness of PdM models up to 54X, which outperforms the state-of-the-art defense by providing 3X more defense against adversarial attacks.

## References

[1] Abomhara, M., Køien, G.M., 2015. Cyber security and the internet of things: vulnerabilities, threats, intruders and attacks. Journal of Cyber Security and Mobility 4, 65–88.

[2] Academy, J., 2018. Layers Of The Internet Of Things. https://analyticstraining.com/4-layers-of-the-internet-of-things/ [Online; accessed 03-Dec-2018].

[3] Akhtar, N., Mian, A., 2018. Threat of adversarial attacks on deep learning in computer vision: A survey. IEEE Access 6, 14410–14430.

[4] Ayvaz, S., Alpay, K., 2021. Predictive maintenance system for production lines in manufacturing: A machine learning approach using iot data in real-time. Expert Systems with Applications 173, 114598.

[5] Backes, J., Bolignano, P., Cook, B., Gacek, A., Luckow, K.S., Rungta, N., Schaef, M., Schlesinger, C., Tanash, R., Varming, C., et al., 2019. One-click formal methods. IEEE Software 36, 61–65.

[6] Basak, A., Rathore, P., Nistala, S.H., Srinivas, S., Runkana, V., 2021. Universal adversarial attack on deep learning based prognostics, in: 2021 20th IEEE International Conference on Machine Learning and Applications (ICMLA), IEEE. pp. 23–29.

[7] Biggio, B., Nelson, B., Laskov, P., 2011. Support vector machines under adversarial label noise, in: Asian conference on machine learning, pp. 97–112.

[8] Chen, J., Jing, H., Chang, Y., Liu, Q., 2019. Gated recurrent unit based recurrent neural network for remaining useful life prediction of nonlinear deterioration process. Reliability Engineering & System Safety 185, 372–382.

[9] Das, N., Shanbhogue, M., Chen, S.T., Hohman, F., Chen, L., Kounavis, M.E., Chau, D.H., 2017. Keeping the bad guys out: Protecting and vaccinating deep learning with jpeg compression. arXiv preprint arXiv:1705.02900 .

[10] Del Pozo, S.M., Standaert, F.X., Kamel, D., Moradi, A., 2015. Side-channel attacks from static power: When should we care?, in: Proceedings of the 2015 Design, Automation & Test in Europe Conference & Exhibition, EDA Consortium. pp. 145–150.

[11] Ding, D., Han, Q.L., Xiang, Y., Ge, X., Zhang, X.M., 2018. A survey on security control and attack detection for industrial cyber-physical systems. Neurocomputing 275, 1674–1683.

[12] Duo, W., Zhou, M., Abusorrah, A., 2022. A survey of cyber attacks on cyber physical systems: Recent advances and challenges. IEEE/CAA Journal of Automatica Sinica 9, 784–800.

[13] Gao, J., Wang, B., Lin, Z., Xu, W., Qi, Y., 2017. Deepcloak: Masking deep neural network models for robustness against adversarial samples. arXiv preprint arXiv:1702.06763 .

[14] Giannetsos, T., Dimitriou, T., 2013. Spy-sense: spyware tool for executing stealthy exploits against sensor networks, in: Proceedings of the 2nd ACM workshop on Hot topics on wireless network security and privacy, ACM. pp. 7–12.

[15] Giordano, D., Giobergia, F., Pastor, E., La Macchia, A., Cerquitelli, T., Baralis, E., Mellia, M., Tricarico, D., 2022. Data-driven strategies for predictive maintenance: Lesson learned from an automotive use case. Computers in Industry 134, 103554.

[16] Goodfellow, I.J., Shlens, J., Szegedy, C., 2014. Explaining and harnessing adversarial examples. arXiv preprint arXiv:1412.6572 .

[17] Guesmi, A., Alouani, I., Khasawneh, K.N., Baklouti, M., Frikha, T., Abid, M., Abu-Ghazaleh, N., 2021. Defensive approximation: securing cnns using approximate computing, in: Proceedings of the 26th ACM International Conference on Architectural Support for Programming Languages and Operating Systems, pp. 990–1003.

[18] Gungor, O., Rosing, T., Aksanli, B., 2022. Stewart: Stacking ensemble for white-box adversarial attacks towards more resilient data-driven predictive maintenance. Computers in Industry 140, 103660.

[19] Hussain, S., Neekhara, P., Dubnov, S., McAuley, J., Koushanfar, F., 2021. {WaveGuard}: Understanding and mitigating audio adversarial examples, in: 30th USENIX Security Symposium (USENIX Security 21), pp. 2273–2290.

[20] Jiang, Y., Wu, S., Yang, H., Luo, H., Chen, Z., Yin, S., Kaynak, O., 2022. Secure data transmission and trustworthiness judgement approaches against cyber-physical attacks in an integrated data-driven framework. IEEE Transactions on Systems, Man, and Cybernetics: Systems .

[21] Jiang, Y., Yin, S., Kaynak, O., 2020. Performance supervised plant-wide process monitoring in industry 4.0: A roadmap. IEEE Open Journal of the Industrial Electronics Society 2, 21–35.

[22] Khalil, R.A., Saeed, N., Masood, M., Fard, Y.M., Alouini, M.S., Al-Naffouri, T.Y., 2021. Deep learning in the industrial internet of things: Potentials, challenges, and emerging applications. IEEE Internet of Things Journal .

[23] Kumar, R., Goyal, R., 2019. On cloud security requirements, threats, vulnerabilities and countermeasures: A survey. Computer Science Review 33, 1–48.

[24] Kurakin, A., Boneh, D., Tramèr, F., Goodfellow, I., Papernot, N., McDaniel, P., 2018. Ensemble adversarial training: Attacks and defenses .

[25] Kurakin, A., Goodfellow, I., Bengio, S., 2016. Adversarial examples in the physical world. arXiv preprint arXiv:1607.02533 .

[26] Lee Neubecker, 2019. Could a sonic weapon have caused the two recent boeing 737 max 8 crashes? URL: https://leeneubecker.com/sonic-weapon-attack-boeing/. [Online; accessed 22-March-2019].

[27] Li, J., Li, X., He, D., 2019. A directed acyclic graph network combined with cnn and lstm for remaining useful life prediction. IEEE Access 7, 75464–75475.

[28] Liao, F., Liang, M., Dong, Y., Pang, T., Hu, X., Zhu, J., 2018. Defense against adversarial attacks using high-level representation guided denoiser, in: Proceedings of the IEEE Conference on Computer Vision and Pattern Recognition, pp. 1778–1787.

[29] Liu, Y., Chen, X., Liu, C., Song, D., 2016. Delving into transferable adversarial examples and black-box attacks. arXiv preprint arXiv:1611.02770 .

[30] Madry, A., Makelov, A., Schmidt, L., Tsipras, D., Vladu, A., 2017. Towards deep learning models resistant to adversarial attacks. arXiv preprint arXiv:1706.06083 .

[31] Maiti, A., Jadliwala, M., He, J., Bilogrevic, I., 2015. (smart) watch your taps: side-channel keystroke inference attacks using smartwatches, in: Proceedings of the 2015 ACM International Symposium on Wearable Computers, ACM. pp. 27–30.

[32] Meng, D., Chen, H., 2017. Magnet: a two-pronged defense against adversarial examples, in: Proceedings of the 2017 ACM SIGSAC Conference on Computer and Communications Security, ACM. pp. 135–147.

[33] Mode, G.R., Anuarul Hoque, K., 2020. Crafting adversarial examples for deep learning based prognostics, in: 2020 19th IEEE International Conference on Machine Learning and Applications (ICMLA), pp. 467–472. doi:10.1109/ICMLA51294.2020.00079.

[34] Mode, G.R., Calyam, P., Hoque, K.A., 2020. Impact of false data injection attacks on deep learning enabled predictive analytics, in: NOMS 2020-2020 IEEE/IFIP Network Operations and Management Symposium, IEEE. pp. 1–7.

[35] Mode, G.R., Hoque, K.A., 2020. Adversarial examples in deep learning for multivariate time series regression. arXiv preprint arXiv:2009.11911 .

[36] Nordal, H., El-Thalji, I., 2021. Modeling a predictive maintenance management architecture to meet industry 4.0 requirements: A case study. Systems Engineering 24, 34–50.

[37] Papernot, N., McDaniel, P., Goodfellow, I., Jha, S., Celik, Z.B., Swami, A., 2017. Practical black-box attacks against machine learning, in: Proceedings of the 2017 ACM on Asia conference on computer and communications security, ACM. pp. 506–519.

[38] Papernot, N., McDaniel, P., Wu, X., Jha, S., Swami, A., 2016. Distillation as a defense to adversarial perturbations against deep neural networks, in: 2016 IEEE Symposium on Security and Privacy (SP), IEEE. pp. 582–597.

[39] Park, K., Choi, Y., Choi, W.J., Ryu, H.Y., Kim, H., 2020. Lstm-based battery remaining useful life prediction with multi-channel charging profiles. IEEE Access 8, 20786–20798.

[40] Qiu, S., Liu, Q., Zhou, S., Wu, C., 2019. Review of artificial intelligence adversarial attack and defense technologies. Applied Sciences 9, 909.

[41] Rahman, M.A., Mohsenian-Rad, H., 2012. False data injection attacks

with incomplete information against smart power grids, in: 2012 IEEE Global Communications Conference (GLOBECOM), IEEE. pp. 3153–3158.

[42] Ramasso, E., Saxena, A., 2014. Performance benchmarking and analysis of prognostic methods for cmapss datasets. International Journal of Prognostics and Health Management 5, 1–15.

[43] Ran, Y., Zhou, X., Lin, P., Wen, Y., Deng, R., 2019. A survey of predictive maintenance: Systems, purposes and approaches. arXiv preprint arXiv:1912.07383 .

[44] Ren, L., Cheng, X., Wang, X., Cui, J., Zhang, L., 2019. Multi-scale dense gate recurrent unit networks for bearing remaining useful life prediction. Future Generation Computer Systems 94, 601–609.

[45] Sadeghi, A.R., Wachsmann, C., Waidner, M., 2015. Security and privacy challenges in industrial internet of things, in: 2015 52nd ACM/EDAC/IEEE Design Automation Conference (DAC), IEEE. pp. 1–6.

[46] Samangouei, P., Kabkab, M., Chellappa, R., 2018. Defense-gan: Protecting classifiers against adversarial attacks using generative models. arXiv preprint arXiv:1805.06605 .

[47] Sarker, I.H., Khan, A.I., Abushark, Y.B., Alsolami, F., 2022. Internet of things (iot) security intelligence: a comprehensive overview, machine learning solutions and research directions. Mobile Networks and Applications , 1–17.

[48] Serradilla, O., Zugasti, E., Rodriguez, J., Zurutuza, U., 2022. Deep learning models for predictive maintenance: a survey, comparison, challenges and prospects. Applied Intelligence , 1–31.

[49] Serradilla, O., Zugasti, E., Zurutuza, U., 2020. Deep learning models for predictive maintenance: a survey, comparison, challenges and prospect. arXiv preprint arXiv:2010.03207 .

[50] Shao, Z., Wu, Z., Huang, M., 2021. Advexpander: Generating natural language adversarial examples by expanding text. IEEE/ACM Transactions on Audio, Speech, and Language Processing 30, 1184–1196.

[51] Sikder, A.K., Petracca, G., Aksu, H., Jaeger, T., Uluagac, A.S., 2018. A survey on sensor-based threats to internet-of-things (IoT) devices and applications. arXiv preprint arXiv:1802.02041 .

[52] Subramanian, V., Uluagac, S., Cam, H., Beyah, R., 2013. Examining the characteristics and implications of sensor side channels, in: 2013 IEEE International Conference on Communications (ICC), IEEE. pp. 2205–2210.

[53] Szegedy, C., Zaremba, W., Sutskever, I., Bruna, J., Erhan, D., Goodfellow, I., Fergus, R., 2013. Intriguing properties of neural networks. arXiv preprint arXiv:1312.6199 .

[54] Tippenhauer, N.O., Pöpper, C., Rasmussen, K.B., Capkun, S., 2011. On the requirements for successful gps spoofing attacks, in: Proceedings of the 18th ACM conference on Computer and communications security, ACM. pp. 75–86.

[55] Tramèr, F., Kurakin, A., Papernot, N., Goodfellow, I., Boneh, D., McDaniel, P., 2017. Ensemble adversarial training: Attacks and defenses. arXiv preprint arXiv:1705.07204 .

[56] Tsipras, D., Santurkar, S., Engstrom, L., Turner, A., Madry, A., 2018. Robustness may be at odds with accuracy. arXiv preprint arXiv:1805.12152 .

[57] US Department of Homeland Security CISA Cyber + Infrastructure, 2019. CAN bus network implementation in avionics. URL: https://www.us-cert.gov/ics/alerts/ics-alert-19-211-01. [Online; accessed 13-September-2019].

[58] Wang, Y., Zhao, Y., Addepalli, S., 2020. Remaining useful life prediction using deep learning approaches: A review. Procedia Manufacturing 49, 81–88.

[59] Wen, Y., Rahman, M.F., Xu, H., Tseng, T.L.B., 2022. Recent advances and trends of predictive maintenance from data-driven machine prognostics perspective. Measurement 187, 110276.

[60] Xie, C., Wang, J., Zhang, Z., Zhou, Y., Xie, L., Yuille, A., 2017. Adversarial examples for semantic segmentation and object detection, in: Proceedings of the IEEE International Conference on Computer Vision, pp. 1369–1378.

[61] Xu, H., Ma, Y., Liu, H.C., Deb, D., Liu, H., Tang, J.L., Jain, A.K., 2020. Adversarial attacks and defenses in images, graphs and text: A review. International Journal of Automation and Computing 17, 151–178.

[62] Xu, W., Evans, D., Qi, Y., 2017. Feature squeezing: Detecting adversarial examples in deep neural networks. arXiv preprint arXiv:1704.01155 .

[63] Yang, B., Liu, R., Zio, E., 2019. Remaining useful life prediction based on a double-convolutional neural network architecture. IEEE Transactions on Industrial Electronics 66, 9521–9530.

[64] Zhang, L., Lin, J., Liu, B., Zhang, Z., Yan, X., Wei, M., 2019a. A review on deep learning applications in prognostics and health management. IEEE Access 7, 162415–162438.

[65] Zhang, X., Dong, Y., Wen, L., Lu, F., Li, W., 2019b. Remaining useful life estimation based on a new convolutional and recurrent neural network, in: 2019 IEEE 15th International Conference on Automation Science and Engineering (CASE), IEEE. pp. 317–322.